\documentclass[10pt]{article}
\usepackage{amsmath}
\usepackage{amsfonts}
\usepackage{amssymb}
\usepackage{xcolor}
\usepackage{color}
\usepackage[english]{babel}
\usepackage{graphics}
\usepackage{graphicx}
\begin{document}
\title{Possible Emission of Cosmic $X$-- and $\gamma$--rays \\
by Unstable Particles at Late Times}
\author{K. Urbanowski\footnote{e--mail: K.Urbanowski@if.uz.zgora.pl},
K. Raczy\'{n}ska\\
Institute of Physics,
University of Zielona Gora,  \\
ul. Prof. Z. Szafrana 4a,
65-516 Zielona Gora,
Poland}
\maketitle

\begin{abstract}
We find that  charged unstable particles
as well as  neutral unstable
particles with non--zero magnetic moment which
live sufficiently long  may emit electromagnetic
radiation. This new mechanism is connected
with the properties of unstable particles at the
post exponential time region. Analyzing the
transition times region between exponential
and non-expo\-nen\-tial form of the survival
amplitude it  is found that the instantaneous
energy of the unstable particle can take very
large values, much larger than the energy of
this state for times from the exponential
time region. Based on the results obtained
for the  model considered, it is shown that this
purely quantum  mechanical effect may  be
responsible for causing unstable particles
to emit electromagnetic--, $X$-- or
$\gamma$--rays  at some time intervals from  the
transition time regions.
\end{abstract}

\section{Introduction}
Not all astrophysical mechanisms of the  emission of
electromagnetic radiation including $X$-- and $\gamma$-- rays
coming from the space  are clear.
Typical physical processes in which cosmic microwave and other
electromagnetic radiation, $X$--, or $\gamma$-rays are generated
have  purely electromagnetic nature
(an acceleration of charged particles, inverse Compton scattering, etc.),
or have the nature of nuclear and particle physics reactions
(particle--antiparticle annihilation, nuclear fusion and  fission,
nuclear or particle decay).
The knowledge of these processes is not sufficient for explaining
all mechanisms  driving the emission  from
some galactic and extragalactic $X$-- or  $\gamma$--rays sources, e.g.
the mechanism that generates $\gamma$--ray emission of
the so-called "Fermi bubbles" remains controversial,
the mechanisms which drive the high energy emission from blazars is still
poorly understood, etc. (see eg. \cite{Holder,Anchordoqui,Gehrels-1,Lipari}).
Similar problems can be encountered when trying to explain the mechanism of radiation
of some cosmic radio sources: Origin of some radio bursts and  many of other sources
is still unknown (see, eg.  \cite{science-1,science-2}) and the radiation
mechanism is unclear and, at the best, insufficiently clear.
Astrophysical processes are the source of not only electromagnetic,
$X$- or $\gamma$-rays but also
a huge number of elementary particles including unstable particles
of very high energies (see eg. \cite{Lipari}).
The numbers of created unstable particles during these processes are
so large that many  of them can survive up to times $t$ at which the
survival probability depending on $t$
transforms from  the exponential form into the inverse power--like form.
It appears that at this time region a new quantum effect is observed:
A very rapid fluctuations of the  instantaneous energy of unstable
particles  take place. These fluctuations of
he instantaneous energy should manifest themselves as fluctuations of the velocity
of the particle. We show that this effect may cause unstable particles to emit
electromagnetic radiation of a very wide  spectrum: from radio--
up to ultra--high frequencies $\nu$ including  $X$--rays and $\gamma$--rays.

To make the paper easily understandable we start in Sec. 2 with a brief
introduction into  the problem
of the late time behavior of unstable states. In Sec. 3 late time
properties of the a energy of  unstable states  are analyzed. In Sec. 4
observable effects are discussed: The emission of electromagnetic
radiation by unstable particles created
in astrophysical processes.
Final Section provides a short summary and
suggestions where to look for signs of the effect
described in Sec. 4.

\section{Late time properties of unstable states }

Searching for the properties of unstable states  one usually analyzes
their decay law, i. e. their survival probability:
If $|\phi\rangle$ is an initial unstable
state then the survival probability, ${\cal P}(t)$, equals
${\cal P}(t) = |a(t)|^{2}$, where
$a(t)$ is the survival amplitude, $a(t) = \langle \phi|\phi;t\rangle$,
and $|\phi;t\rangle = e^{\textstyle{-itH}}\,|\phi\rangle$, $H$ is the
total Hamiltonian of the system under considerations,
$|\phi \rangle, |\phi;t\rangle
\in {\cal H}$ and ${\cal H}$ is the Hilbert space of states of
the considered system.
The spectrum, $\sigma(H)$, of $H$ is assumed to be bounded from below,
$\sigma(H) =[E_{min},\infty)$ and $E_{min} > -\infty$.
Studying the late time properties of unstable states
it is convenient to use the integral representation of
$a(t)$ as the Fourier transform
of the energy distribution function,
$\omega(E)$,
\begin{equation}
a(t) = \int\,\omega(E)\,e^{\textstyle{-itE}}\,dE, \label{a(t)-rho}
\end{equation}
with $\omega (E) \geq 0$ and $\omega (E) = 0$ for $E < E_{min}$ \cite{Khalfin,Fonda,Chiu,Sluis,muga-1,calderon-2,nowakowski,urbanowski-2-2009,urbanowski-1-2009,Krauss}.
In the case of quasi--stationary (metastable) states
it is useful to express  $a(t)$  in the
following form \cite{urbanowski-2-2009,urbanowski-1-2009}, $
a(t) = a_{exp}(t) + a_{lt}(t)$,
where $a_{exp}(t)$ is the exponential part of $a(t)$, that is $a_{exp}(t) =
N\,\exp\,[{-it(E_{\phi}^{0} - \frac{i}{2}\,{\it\Gamma}_{\phi}^{0})}]$,
($E_{\phi}^{0}$ is the energy of the system in the state $|\phi\rangle$
measured at the canonical decay times,
i.e. when ${\cal P}_{\phi}(t)$ has the exponential form,
${\it\Gamma}_{\phi}^{0}$ is the decay width, $N$ is the normalization
constant), and $a_{lt}(t)$ is the
late time non--exponential part of $a(t)$.

\begin{figure}[h!]
\begin{center}
\includegraphics[width=68mm]{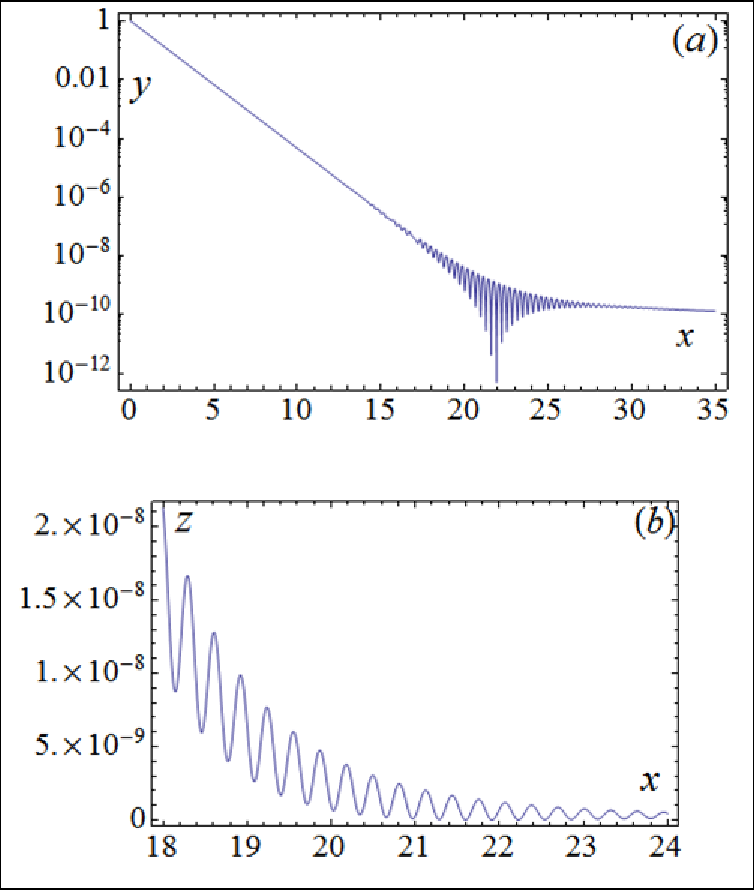}
\caption{Axes:
$x =t / \tau_{\phi}, y = {\cal P}(t)$ --- the logarithmic scale, $z = {\cal P}(t)$.
(a) The general, typical  form of the decay curve ${\cal P}(t)$.
(b) An enlarged part of (a) showing a typical behavior of the survival probability
${\cal P}(t)$ at the transition times region when $t \sim T$.}
  \label{fa}
\end{center}
\end{figure}

From the literature it is known  that the characteristic  feature of
survival probabilities ${\cal P}(t)$ is the presence of sharp and frequent fluctuations at
the transition times region, when contributions from $|a_{exp}(t)|^{\,2}$ and
$|a_{lt}(t)|^{\,2}$ into ${\cal P}(t)$ are comparable
(see, eg. \cite{Fonda,Sluis,muga-1,calderon-2,nowakowski}), and that the
amplitude  $a_{lt}(t)$ and thus the probability ${\cal P}(t)$ exhibits inverse
power--law behavior at the late time region for times $t$ much later than the
crossover time $T$.
(This effect was confirmed experimentally not long ago \cite{rothe}).
The crossover time $T$
can be found by solving the following equation, $|a_{exp}(t)|^{\,2} = |a_{lt}(t)|^{\,2}$.
In general $T \gg \tau_{\phi}$, where
$\tau_{\phi} = 1/{\it\Gamma}^{0}_{\phi}$ is the live--time of $\phi$.
Formulae for $T$ depend on the model considered (i.e. on $\omega(E)$) in general
(see, eg.  \cite{Chiu,Sluis,muga-1,urbanowski-2-2009,urbanowski-1-2009,Krauss}).
The standard form of the decay curve, that is the form of the probability ${\cal P}(t)$
as a function of time $t$ is presented in Fig. (\ref{fa}).
In this Figure the calculations were performed using
the  Breit--Wigner   energy distribution function,
$\omega ({E}) \equiv \omega_{BW}(E)$, where
\begin{equation}
\omega_{BW}(E) \stackrel{\rm def}{=}
 \frac{N}{2\pi}  {\it\Theta} ( E - E_{min})
\frac{{\it\Gamma}_{\phi}^{0}}{( E - E_{\phi}^{0})^{2} +
(\frac{{\it\Gamma}_{\phi}^{0}}{2})^{2}}, \label{omega-BW}
\end{equation}
and $\it\Theta ({E})$ is the unit step function. In Fig. (\ref{fa})
calculations were performed for
$({E_{\phi}^{0}} - E_{min}) / {{\it\Gamma}_{\phi}^{0}} = 20$.

Deviations form the exponential decay law visible in Fig. (\ref{fa}) are
caused by the regeneration process {\cite{Fonda,Fonda-2,muga-3}.
A certain fixed proportion between rates of  decay  and  regeneration
processes does not change at canonical decay times, so there is a kind of a balance
between these processes at this time region. This  balance
is broken at transition times and later. Oscillations of the decay law seen
in the transition times region are a reflection of this fact.
Time intervals around local maxima of the decay curve
presented in the panel (b) of Fig. (\ref{fa})
are places where the regeneration process begins to  dominate temporarily  and
the decay process slows down. On the other hand, times close to the local
minima fix places where the regeneration rate
is minimal and the decay process accelerates.

Note that the survival amplitude $a(t)$ obtained within quantum mechanics
share with the amplitude $a(t)$ obtained as a result of investigations
on relativistic quantum field theory models, the property (\ref{a(t)-rho})
of being the Fourier transform of a positive definite function $\omega(E)$
with a limited from below support (see, eg.
\cite{Chiu,testa,giacosa-1,giacosa-2,giacosa-3}). This means that  effects
connected with the long time behavior at
$t \sim T$ and $t \gg T$  of the survival probability $ {\cal P}(t)$
should take place in the both cases: When they are considered
at the level of quantum mechanical processes as well as at
the level of the processes that require  quantum field theory to describe them.

\section{Energy of unstable states at late times}
It is commonly known that the information about the
decay rate, $\it\Gamma_{\phi}$, of the
unstable state $|\phi\rangle$ under considerations can be extracted
from the survival amplitude $a(t)$. In general not
only  $\it\Gamma_{\phi}$ but also
the instantaneous energy ${\cal E}_{\phi}(t)$ of
an unstable state $|\phi\rangle$ can be calculated
using $a(t)$ \cite{urbanowski-2-2009,urbanowski-1-2009}.
In the considered case,
${\cal E}_{\phi}(t)$ can be found using
the effective Hamiltonian,  $h_{\phi}(t)$,
governing the time evolution in
an one--dimensional subspace of states
spanned  by vector $|\phi\rangle$,
\cite{urbanowski-2-2009,urbanowski-1-2009}:
\begin{eqnarray}
h_{\phi}(t)  &=&  \frac{i}{a(t)}\,\frac{\partial a(t)}{\partial t} \label{h1} \\
&\equiv&   \frac{\langle \phi|H|\phi;t\rangle}{\langle \phi|\phi;t\rangle}.\label{h1a}
\end{eqnarray}
The instantaneous energy ${\cal E}_{\phi}(t)$ of
the system in the state $|\phi\rangle$
is the real part of $h_{\phi}(t)$,
${\cal E}_{\phi}(t) = \Re \,(h_{\phi}(t))$. The imaginary
part of $h_{\phi}(t)$ defines the
instantaneous decay rate $\it\Gamma_{\phi}(t)$, ${\it\Gamma}_{\phi}
\equiv {\it\Gamma}_{\phi}(t) = -2\,\Im\,(h_{\phi}(t))$,
\cite{pra,urbanowski-2-2009,urbanowski-1-2009}.

There is ${\cal E}_{\phi}(t)= E_{\phi}^{0}$ and ${\it\Gamma}_{\phi}(t) =
{\it\Gamma}_{\phi}^{0}$ at the canonical decay
times  (see, eg., \cite{pra})  and at asymptotically late
times (see \cite{urbanowski-1-2009,urbanowski-2-2009,urbanowski-2011}),
\begin{eqnarray}
{\cal E}_{\phi}(t) &\simeq& E_{min} + \frac{c_{2}}{t^{2}}
\,+\,\frac{c_{4}}{t^{4}} \ldots, \;\;\;({\rm for}
\;\;t \gg T), \label{E(t)} \\
\it\Gamma_{\phi}(t)  &\simeq& \frac{c_{1}}{t}
+ \frac{c_{3}}{t^{3}} + \ldots \;\;\;({\rm for}
\;\;t \gg T), \label{g(t)}
\end{eqnarray}
where $c_{i} = c_{i}^{\ast},\;i=1,2,\ldots$, ($c_{1} > 0$ and
the sign of $c_{i}$ for $i \geq 2$ depends on the model considered),
so $\lim_{t \rightarrow \infty}\, {\cal E}_{\phi}(t) = E_{min}$
and $\lim_{t \rightarrow \infty}\, {\it\Gamma}_{\phi}(t) = 0$
\cite{urbanowski-1-2009,urbanowski-2-2009,urbanowski-2011}.
Results (\ref{E(t)}) and (\ref{g(t)}) are rigorous.
The basic physical factor forcing
the  amplitude $a(t)$ to exhibit inverse power law behavior at $t \gg T$ is the
boundedness from below of  $\sigma (H)$. This means that if this condition is satisfied
and $\int _{-\infty}^{+\infty}\omega(E)\,dE\,<\,\infty$, then all  properties
of $a(t)$, including the  form of the time--dependence at  $t \gg T$, are the
mathematical consequence of them both.
The same applies by (\ref{h1}) to the properties of $h_{\phi}(t)$ and concerns
the asymptotic form of $h_{\phi}(t)$ and
thus of ${\cal E}_{\phi}(t)$ and $\it\Gamma_{\phi}(t)$ at $t \gg T$.

The  sharp and frequent of fluctuations of ${\cal P}(t)$ at
the transition times region (see Fig. (\ref{fa})) are a consequence of a similar behavior
of real and imaginary parts of the amplitude $a(t)$ at this time region.
Therefore the derivatives of  $a(t)$ may reach extremely large negative and positive
values for some times from the transition time region and the
modulus of these derivatives is much larger than the modulus
of $a(t)$, which is very small for these times. This
means that at this time region the real part of $h_{\phi}(t)$
which is expressed by the relation (\ref{h1}), i. e. by a large
derivative of $a(t)$  divided by a very small $a(t)$, can reach
values much larger than the energy ${ E}_{\phi}^{0}$ of
the  unstable state measured at the canonical decay times.
Using relations (\ref{a(t)-rho}), (\ref{h1}) and assuming the form of $\omega(E)$ and
performing all necessary calculations numerically one can see
how this mechanism work. A typical behavior  of the instantaneous energy ${\cal E}_{\phi}(t)$
at the transition time region is presented in Figs (\ref{f1}) and (\ref{f2}).
In these figures
the calculations were performed for  the  Breit--Wigner
energy distribution function (\ref{omega-BW}).
\begin{figure}[h!]
\begin{center}
\includegraphics[width=70mm]{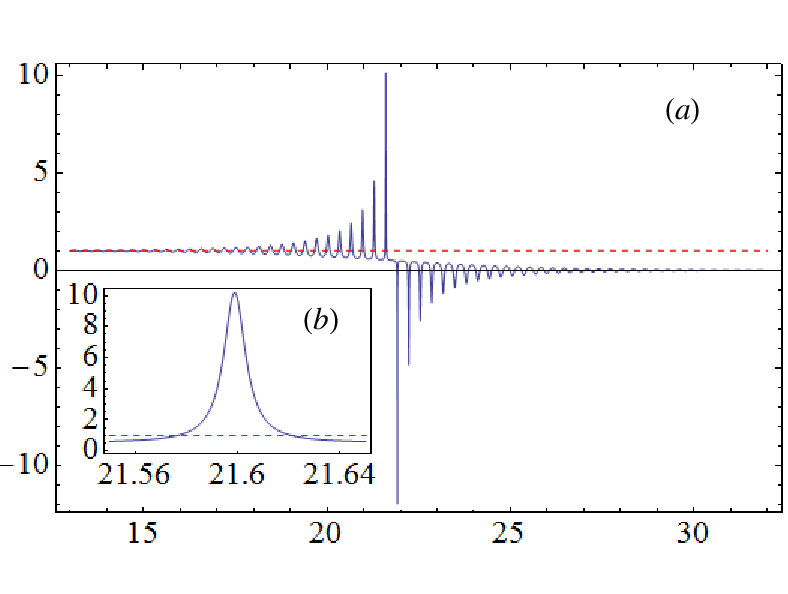}
\caption{Axes: $y= ({\cal E}_{\phi}(t) - E_{min}) / (E_{\phi}^{0} - E_{min})$, $x =t / \tau_{\phi}$.
 The dashed line denotes the  straight line $y=1$.
(a) The instantaneous energy ${\cal E}_{\phi}(t)$
  in the transitions time region:  The case
 $({E_{\phi}^{0}} - E_{min}) / {{\it\Gamma}_{\phi}^{0}} = 20$.
 (b) Enlarged part of (a): The highest maximum
 of $({\cal E}_{\phi}(t) - E_{min}) / (E_{\phi}^{0} - E_{min})  $
in the transition times region.}
  \label{f1}

{\includegraphics[width=70mm]{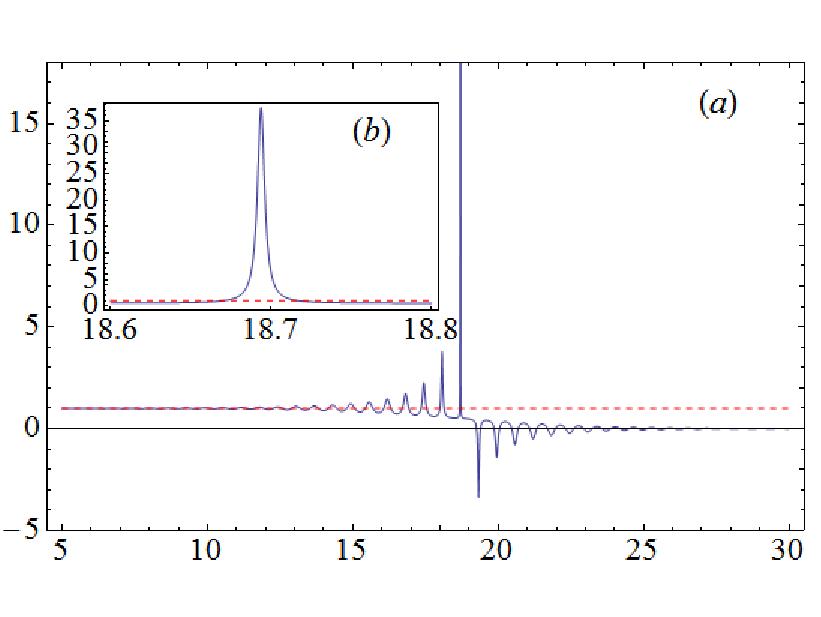}}
\caption{The same as in Fig (1) for
$({E_{\phi}^{0}} - E_{min}) / {{\it\Gamma}_{\phi}^{0}} = 10$. }
  \label{f2}
\end{center}

\end{figure}

From (\ref{h1a}) it follows that the effective Hamiltonian $h_{\phi}(t)$ is
the so-called weak value of $H$ \cite{aharonov-1,aharonov-2,aharonov-3,hosoya}.
Considering $h_{\phi}(t)$ as a weak value
the behavior of ${\cal E}_{\phi}(t)$ at transition times region
and extremely large values reached by ${\cal E}_{\phi}(t)$ at some times
there that can be seen in Figs (\ref{f1}) and (\ref{f2})
are not extraordinary effects.
Properties of this kind are typical for many weak values of physical
quantities \cite{aharonov-1,aharonov-2,aharonov-3,hosoya}.
What is more experiments have verified  aspects of the theory
of weak values (see, eg. \cite{aharonov-3,hosoya,starling,dixon,susa,shomroni}).

It seems that one should observe a picture
presented in Figs (\ref{f1}), (\ref{f2}), or similar one, e.g.
after performing a suitable modification of the experiment described in \cite{rothe}.
This modification should allow one to register not only the presence of the photons
emitted by excited molecules but also  energies of these photons (i.e. frequencies
of the registered radiation). Analogous modifications
of possible experiments based on the effects analyzed
in \cite{muga-2} and proposed there seems to make such the observation possible.

More detailed numerical analysis of ${\cal E}_{\phi}(t)$
and ${\cal P}(t)$ within the model considered
shows that local maxima of
$({\cal E}_{\phi}(t) - E_{min}) / (E_{\phi}^{0} - E_{min})$ correspond with the local minima
of the survival probability ${\cal P}(t)$ (see Fig. (\ref{f2a})).
It is just as one would expect:
The higher the energy ${\cal E}_{\phi}(t)$, i. e.,
the greater the difference
$({\cal E}_{\phi}(t) - E_{min})$ the higher the probability
of a decay (i. e., the survival probability less).
One meets an analogous effect
in the case of the local minima of
$({\cal E}_{\phi}(t) - E_{min}) / (E_{\phi}^{0} - E_{min})$:
They correspond with the local maxima of the survival probability.
There is a simple and obvious interpretation
of this effect: The difference
$({\cal E}_{\phi}(t) - E_{min})$ smaller the decay process
slower and the regeneration process faster.

\begin{figure}
\begin{center}
\includegraphics[width=70mm]{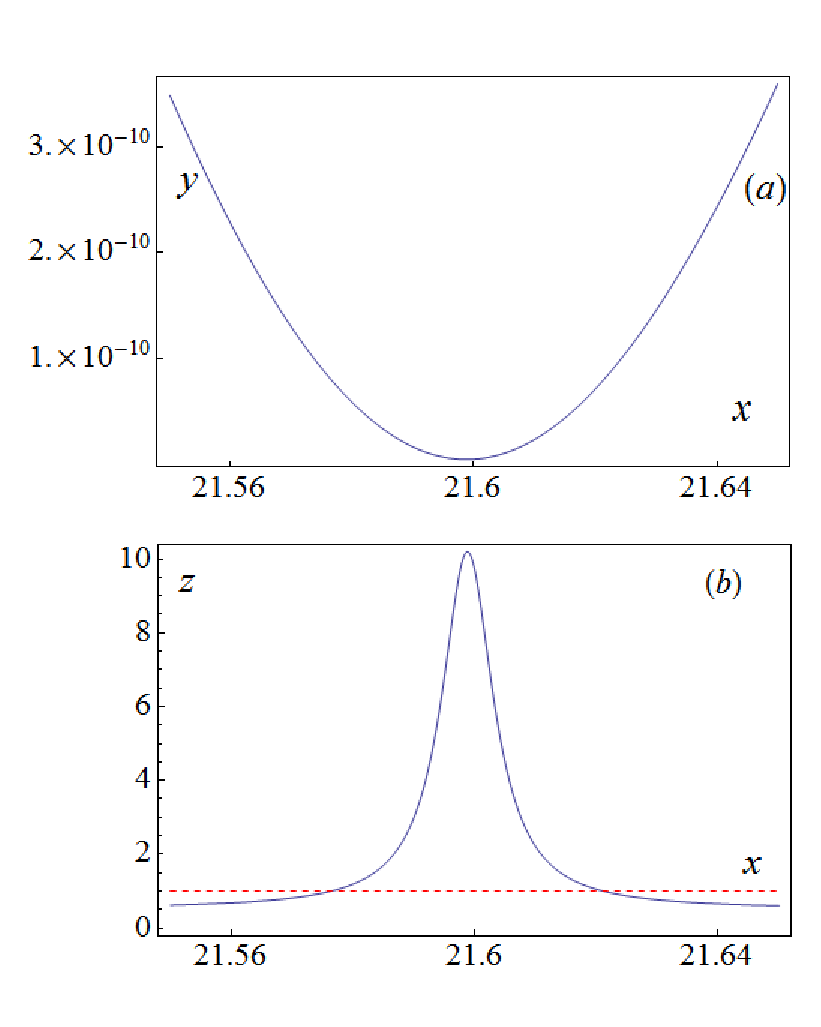}
\caption{Axes:
$x =t / \tau_{\phi}, y = {\cal P}(t)$, $z =
({\cal E}_{\phi}(t) - E_{min}) / (E_{\phi}^{0} - E_{min})$.
The case  $({E_{\phi}^{0}} - E_{min}) / {{\it\Gamma}_{\phi}^{0}} = 20$.
(a) The local minimum of ${\cal P}(t)$ corresponding with
the highest local maximum of the ratio
$({\cal E}_{\phi}(t) - E_{min}) / (E_{\phi}^{0} - E_{min})$.
(b) The highest local maximum of
$ ({\cal E}_{\phi}(t) - E_{min}) / (E_{\phi}^{0} - E_{min})$.}
  \label{f2a}
\end{center}
\end{figure}

From the results presented in Figs (\ref{f1}) and (\ref{f2})
one can see that the ratio,
$({\cal E}_{\phi}(t) - E_{min}) / (E_{\phi}^{0} - E_{min})$,
takes negative values for some times at
transition times region. This does not mean that the instantaneous
energy ${\cal E}_{\phi}(t)$ of the unstable particle
takes  negative negative values at these times $t$. The negative
values of $({\cal E}_{\phi}(t) - E_{min}) / (E_{\phi}^{0} - E_{min})$
mean that ${\cal E}_{\phi}(t) > 0 $ becomes smaller than $E_{min}$
at these $t$. The most negative values of this ratio occur in its local minima,
which correspond to local maxima of the survival probability ${\cal P}(t)$
as it was mentioned above. This means that at these times the rate of
the decay process greatly slows down  and even nearly
stops but the rate of the regeneration process becomes extremely fast.

\section{Observable effects}

Note that from the point of view of a frame of reference  in which the time evolution
of the unstable system was   calculated the Rothe experiment as well as the
picture presented in Figs (\ref{f1}), (\ref{f2}) refer to the rest coordinate system of the
unstable system considered. Astrophysical sources of unstable particles emit
them  with relativistic or ultra--relativistic velocities in relation to an external
observer so many of these particles move in space  with ultra high energies.
The question is what effects can be observed by an external observer when the unstable
particle, say $\phi$,   which survived up to the transition times region, $t\sim T$, or longer
is moving with a relativistic velocity in relation to this observer.
The distance $d$ from the source reached by this particle is of order
$d \sim d_{T}$, where $d_{T} = v^{\phi} \cdot T',\, T' = \gamma_{L} \, T$ and
$\gamma_{L} \equiv \gamma_{L} (v^{\phi})= (\sqrt{1-\beta^{2}})^{-1}$, $ \beta = v^{\phi}/c$,
$v^{\phi}$ is the velocity of the particle  $\phi$. (For simplicity we assume that
there is  a frame of reference common for the source and observer
both and that they do not move with respect to this frame of reference).
The
relation (\ref{h1a}) explains why effects of type
(\ref{E(t)}), (\ref{g(t)}) and those one can see in Figs (\ref{f1}), (\ref{f2}) are possible.
In the case of moving particles created in astrophysical processes one should consider the
effect shown in Figs (\ref{f1}), (\ref{f2}) together with   the fact that the particle
gains extremely huge kinetic energy, $W^{\phi}$,
which have
to be conserved. There is $W^{\phi} = m^{0}_{\phi} \ c^{2} \ \gamma_{L}$, where
$m_{\phi}^{0}$ is the rest mass of the particle $\phi$. We have
$m_{\phi}^{0} \ c^{2} \equiv E_{\phi}^{0}$ at canonical decay times and thus
$W^{\phi} \equiv E_{\phi}^{0} \ \gamma_{L}$ at these times.
At this time region $E_{\phi}^{0} = {\cal E}_{\phi}(t)$
but at times $t \gg \tau_{\phi}, \; t \sim T$
we have ${\cal E}_{\phi}(t) \neq E_{\phi}^{0}$.
A general relation between instantaneous energies of
the unstable particle in the rest system and in the system
connected with the moving particle
can be found using a relation between
the survival amplitude, $a^{v\neq0}(t)$,  of a moving unstable particle and
the survival amplitude, $a^{v=0}(t)$, of the particle in
the rest coordinate system of the observer $\cal O$. In such a case
assuming that the
rest system of the particle moves with a velocity $v^{\phi}$ relative
to $\cal O$ one can find within
the relativistic quantum theory that (see eg. \cite{exner})
\begin{equation}
a^{v \neq 0}(t) = a^{v=0}(\frac{t}{\gamma_{L}}).  \label{a-v=a}
\end{equation}
The relation  (\ref{a-v=a}) means that
survival probabilities  ${\cal P}^{v\neq 0}(t)$
and ${\cal P}^{v=0}(t/ \gamma_{L})$ corresponding with the survival amplitudes
$a^{v\neq0}(t)$, $a^{v=0}(t/ \gamma_{L})$
respectively are equal.
This property and thus the relation (\ref{a-v=a})
was tested by numerous experiments.
Now using (\ref{h1}) and (\ref{a-v=a}) it is easy to find that
\begin{equation}
h_{\phi}^{v\neq 0} (t) = \frac{1}{\gamma_{L}}\,h_{\phi}^{v=0}(t/ \gamma_{L}),
\label{h-v}
\end{equation}
where $h_{\phi}^{v\neq 0} (t)$ is obtained by inserting into (\ref{h1})
the survival amplitude $a^{v \neq 0}(t)$, and so on.
From this last relation is follows that the instantaneous energy
${\cal E}^{v=0}_{\phi}(t/ \gamma_{L}) = \Re\,(h_{\phi}^{v=0}(t/ \gamma_{L}))$ of the moving
particle $\phi$ measured by the observer ${\cal O}$ equals ${\cal E}^{v=0}_{\phi}(t / \gamma_{L}) =
\gamma_{L}\,{\cal E}^{v\neq 0}_{\phi}(t)$.
Taking into account that ${\cal E}^{v\neq 0}_{\phi}(t) =
\Re\,(h_{\phi}^{v\neq0}(t))$ is the instantaneous
energy measured in the rest system of the particle one can identify it
with the instantaneous energy ${\cal E}_{\phi}(t)$  analyzed in the previous Section.
So in the general case the kinetic energy of the moving particle
$\phi$ having the energy ${\cal E}_{\phi}^{k}$ in its rest system
measured by the observer ${\cal O}$ equals
$W_{k}^{\phi} = \gamma_{L}^{k} \,{\cal E}_{\phi}^{k} $
and here $\gamma_{L}^{k} = \gamma_{L}(v^{\phi}_{k})$.
Similarly there is
$W_{l}^{\phi} = \gamma_{L}^{l} \,{\cal E}_{\phi}^{l}$
for the other particle moving with the velocity $v^{\phi}_{l} \neq v^{\phi}_{k}$
and having the energy ${\cal E}_{\phi}^{l}$.
Now if to assume that we observe the particle $\phi$ at different instants $t_{k} \neq t_{l}$
of time $t$ then we can use the following identification: ${\cal E}_{\phi}^{k(l)} = {\cal E}_{\phi}(t_{k(l)})$.
Of course the  kinetic energies
$W^{\phi}, W^{\phi}_{k}, W^{\phi}_{l}$ of  $\phi$ have to be
the same at the canonical decay times region and  at the transition times
$t_{k},\,t_{l}\, \sim\, T$: $W^{\phi} \equiv W^{\phi}_{k} \equiv W^{\phi}_{l}$, that is there should be
\begin{equation}
W^{\phi} \equiv \gamma_{L}^{k}\,{\cal E}_{\phi}^{k} \,= \, const. \label{W=W}
\end{equation}
From  relation (\ref{W=W}) one can infer
that this is possible only when the changes
of ${\cal E}_{\phi}(t_{k})$ at times $t_{k} \sim T$ are
balanced with suitable changes of $\gamma_{L}^{k} $ (i.e. of  the
velocity $v^{\phi}_  {k}$ of the considered particle).
So, in the case of moving unstable particles, an external observer should detect
rapid fluctuations (changes) of their velocities at
distances $d \sim d_{T}$  from their source. These fluctuations of
the velocities  mean for the observer that the particles are moving
with a nonzero acceleration in this space region, $\dot{v^{\phi}} \neq 0$.
So we can expect that this observer will register electromagnetic radiation
emitted by charged unstable particles, which survived up to times
$t \sim T$, i.e. which reached distances $d \sim d_{T}$ from the source
(see Fig (\ref{f3})).
\begin{figure}[h!]
\begin{center}
\includegraphics[width=80mm]{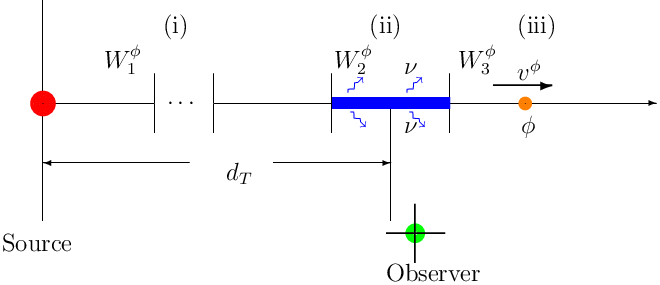} \\
\caption{
Time regions:
(i) Canonical decay, (ii) Transition, (iii) Asymptotically late.
 $W_{i}^{\phi} = \gamma_{L}^{i}\,{\cal E}_{\phi}^{i},\;(i=1,2,3)$ and $W_{i}^{\phi}$ is
 the energy of moving relativistic particle $\phi$ measured by the observer, ${\cal E}_{\phi}^{i} = {\cal E}_{\phi}(t_{i})$, $t_{1} \ll t_{2} \ll t_{3}$,
 $t_{1} \sim \tau_{\phi}$, $t_{2} \sim T$,  $t_{3} \gg T$ and
 $ W^{\phi}_{1}  = E_{\phi}^{0} \gamma_{L}^{1}$,
 $\nu$ is the frequency of the emitted electromagnetic rays.
 }
\label{f3}
\end{center}
\end{figure}
This follows from the Larmor formula and its relativistic generalization, which
state that the  total radiation power, $P$, from the considered charged particle
is proportional to $(\dot{v^{\phi}})^{2}$ (see eg. \cite{Jackson}):
\begin{equation}
P = \frac{1}{6 \pi \epsilon_{0}} \, \frac{ q^{2}  \,
({ \dot{v^{\phi}} })^{2} }{ c^{2} }  \, \gamma_{L}^{6},
\label{dwdt}
\end{equation}
(where $ q $ is the electric charge, $ {\epsilon}_{0} $ -- permittivity for free space),
 and $\dot{v^{\phi}} \neq 0$ implies that there must be $P\neq 0$.
The same conclusion also concerns neutral unstable particles
with non--zero magnetic moment \cite{Jackson,Griffiths}.
One should expect that the spectrum of this radiation will be very wide: From high
radio frequencies, through $X$--rays up to high energy $\gamma$--rays  depending on the scale
of the fluctuations of the instantaneous energy ${\cal E}_{\phi}(t)$ in this space region.

Within  the model defined by
$\omega_{BW}(E)$
the cross--over time $T$ can be found
using the following approximate relation valid for ${E_{\phi}^{0}}/{\it\Gamma_{\phi}^{0}} \gg 1 $,
\cite{urbanowski-2-2009}:
\begin{equation}
 {\it\Gamma}_{\phi}^{0}\,T \equiv \frac{T}{\tau_{\phi}} \;
 \sim\;2\, \ln \,\Big[2 \pi \Big(\frac{E_{\phi}^{0} - E_{min}}{\it\Gamma_{\phi}^{0}}\Big)^{2}\Big],
\label{t-as}
\end{equation}
whereas for the model considered in \cite{Krauss,Chiu} one has
${T}/{ \tau_{\phi}}\; \sim \;5\,\ln \big[({E_{\phi}^{0}} -
E_{min})/{{\it\Gamma}_{\phi}^{0}}\big] $
(see (11) in \cite{Krauss}). Considering a meson $\mu^{\pm}$ as an example
and taking $E_{\phi}^{0} - E_{min} = m_{\mu^{\pm}} - (m_{e}
+ m_{\bar{\nu}_{e}} + m_{\nu_{\mu}}) \simeq 105$ [MeV],
then using (\ref{t-as})  one finds $T = T_{\mu} \sim 165.3\,\tau_{\mu}$.
(The formula (11) from \cite{Krauss} gives  $T_{\mu} \sim 202 \,\tau_{\mu}$).
The distance $d_{T_{\mu}}$ from the source reached by muon, which survived up to the
time $T_{\mu} \simeq 165.3\,\tau_{\mu}$ depends on its kinetic energy $W^{\mu}$ and
equals:  from $d_{T_{\mu}} \simeq 10^{6}$ [m] if  $W^{\mu} = 10^{9}$ [eV], up to
$d_{T_{\mu}} \simeq 0.033$ [pc] if $W^{\mu} = 10^{18}$ [eV].
Similarly, there is for $\pi$--mesons
$E_{\phi}^{0} - E_{min} =m_{\pi^{\pm}} -(m_{\mu^{\pm}}
+ m_{\nu_{\mu}}) \simeq 33.83 $ [MeV], which leads by (\ref{t-as})
to the results: $T_{\pi^{\pm}} \simeq 143 \;\tau_{\pi^{\pm}}$
and $d_{T_{\pi^{\pm}}} 7.9 \simeq  \times 10^{3}$ [m] if  $W^{\pi} = 10^{9}$ [eV]
and $d_{T_{\pi^{\pm}}} \simeq 8 \times 10^{12}$ [m]
$\simeq 53.45$ [au]  if  $W^{\pi} = 10^{18}$ [eV].
For the neutron $E_{\phi}^{0} - E_{min} =m_{n}  -
(m_{p} + m_{e} + m_{\nu_{e}}) \simeq 0.78$ [MeV] and
using (\ref{t-as})
one finds: $T_{n} \simeq 225 \;\tau_{n}$ and $d_{T_{n}} \simeq 130.6 $ [au] if
$W^{n} = 10^{9}$ [eV] and $d_{T_{n}} \simeq 2.05$ [Mpc] if  $W^{n} = 10^{18}$ [eV].

Let us now analyze   Fig (\ref{f1}) in more details.
Coordinates of the highest maximum
in Fig (\ref{f1}) are equal: ($x_{mx}, y_{mx}) = (21.60, 10.27)$. Coordinates of points of
the intersection of this maximum with the straight line $y=1$ are equal: $(x_{1}, y_{1})
= (21.58, 1.0)$ and $(x_{2}, y_{2}) = (21.62, 1.0)$. From these coordinates
one can extract the change $\Delta v^{\phi} = v^{\phi}_{M} - v^{\phi}_{1} $ of
the velocity $v^{\phi}$ of the considered particle and the time interval
$\Delta t = t_{M} - t_{1}$ at which this change occurred (Here $t_{M} = t_{mx}$ and  $v^{\phi}_{1} = v^{\phi}(t_{1})$). Indeed,
using (\ref{W=W}) one finds
\begin{equation}
\gamma_{L}^{1}  = \frac{{\cal E}_{\phi}^{M}}{{\cal E}_{\phi}^{1}} \;
\gamma_{L}^{M}. \label{E2/E1}
\end{equation}
There are $ {\cal E}_{\phi}^{M} = {\cal E}_{\phi}(t_{M})$ and
 ${\cal E}_{\phi}^{1} = {\cal E}_{\phi}(t_{1}) \equiv E^{0}_{\phi}$ in the considered case. This means
that we can replace $\gamma_{L}^{1}$ by $\gamma_{L}$ measured at the canonical decay
times and then taking the value of the ratio ${\cal E}_{\phi}^{M} / E_{\phi}^{0}$
from Fig (\ref{f1}) we can use (\ref{E2/E1}) to calculate $\gamma_{L}^{M}$.
Figs (\ref{f1}), (\ref{f2}) show how the ratio
$({\cal E}_{\phi}(t) - E_{min}) / (E_{\phi}^{0} - E_{min})
\stackrel{\rm def}{=} \kappa (t)$ varies
in time and this $\kappa (t) $ can be easy extracted form these
Figures, eg., for $t=t_{M}$ and for $t=t_{1} \neq t_{M}$.
If one wants to use the relation (\ref{E2/E1}) in order to
calculate $\gamma_{L}^{M}$, one needs the ratio
${\cal E}_{\phi}(t) / E_{\phi}^{0}$ instead of $\kappa (t)$.
Using $\kappa (t)$ it is easy to express ${\cal E}_{\phi}(t) / E_{\phi}^{0}$
in terms of known parameters of unstable particles considered. We have
\begin{equation}
\frac{{\cal E}_{\phi}(t)}{ E_{\phi}^{0}} =
\kappa (t)\,-\,\big(\kappa (t)\,-\,1\big)\,\frac{E_{min}}{E_{\phi}^{0}}. \label{E/E0}
\end{equation}
In the considered case $\kappa (t_{M}) \equiv y_{mx} = 10.27$ which, eg., for the muon gives
${\cal E}_{\mu}^{M}/ E_{\phi}^{0}\simeq 10.21$.
Hence using (\ref{E2/E1})  within the considered model one finds that  for the muon there is
$\gamma_{L}^{1} \simeq 10.21\,\gamma_{L}^{M}$.
Next having $\gamma_{L}^{1} \equiv \gamma_{L}$ and $\gamma_{L}^{M}$ it is easy to
find $\Delta v^{\phi} = v^{\phi}_{M} - v^{\phi}_{1}$. Now using (\ref{dwdt})
 one can estimate the energy $P$ of the electromagnetic radiation emitted in unit of
time by an unstable charged relativistic particle $\phi$
during the time interval $\Delta t$. In other words,
one can find
$\Delta v^{\phi} / \Delta t$ and thus $P  \propto  (\Delta v^{\phi} / \Delta t)^{2}$.
This procedure, formulae (\ref{W=W}), (\ref{E2/E1}) and parameters describing the
highest maximum  in Fig. (\ref{f1})
lead to the following (simplified, very conservative) estimations of the
energies of the electromagnetic radiation emitted by ultra relativistic
muon at the transition times region (in a distance $d \sim d_{T}$ from the source):
$P \sim 4.6$ [eV/s].
Analogously  coordinates of the highest maximum
in Fig (\ref{f2}) are equal: ($x_{mx}, y_{mx}) = (18.69, 37.68)$ and coordinates of points of
the intersection of this maximum with the line $y=1$ are: $(x_{1}, y_{1})
= (18.67, 1.0)$ and $(x_{2}, y_{2}) = (18.72, 1.0)$. This leads to the following estimation:
$P \sim 0.84 $ [keV/s].
Similar estimations of $P$ can be found for neutral
ultra--relativistic unstable particles with non--zero magnetic moment.

The question is where the above described effect may be observed.
Astrophysical and cosmological processes in which  extremely
huge numbers of unstable particles are created
seem to be  a possibility for the above discussed
effect to become manifest.
The fact is that
the probability ${\cal P}_{\phi}(t)
= |a(t)|^{2}$ that an  unstable particle $\phi$ survives
up to time $t \sim T$ is extremely small. Let
${\cal P}_{\phi}(t)$ be  ${ {\cal P}_{\phi}(t)\,\vline }_{\;t \sim T}\;\sim\;10^{-k}$,
where $k \gg 1$, then there is a chance to observe some of
particles $\phi$ survived at $t \sim T$ only if there
is a source creating these particles in ${\cal N}_{\phi}$
number such that  ${{\cal P}_{\phi}(t)\,\vline }_{\;t \sim
T}\;{\cal N}_{\phi} \; \gg \;1 $.
So if  a source exists that creates a flux containing
${\cal N}_{\phi} \;\sim\;10^{\,l} $,
unstable particles and $l \gg k$ then the probability
theory states that the number $N_{surv}$ of unstable particles
$N_{surv} = { {\cal P}_{\phi}(t)\,\vline }_{\;t \sim T}\;
{\cal N}_{\phi} \;\sim\;10^{l - k} \; \gg\;1 $,
has to survive up to time $t \sim T$. Sources creating
such numbers of unstable particles are known from cosmology
and astrophysics: as example of such a source can be considered
processes taking place in galactic  nuclei (galactic cores), inside
stars, etc. According to estimations of the luminosity of some
$\gamma$--rays sources the energy  emitted  by  these sources  can even
reach a value of order $10^{52}$ [erg/s], \cite{Lipari,Letessier,Hinton,Gehrels},
and it is only a part of the total energy produced there. So, if one has a source
emitting energy $10^{50}$ [erg/s] then, eg., an emission of
${\cal N}_{0} \simeq 6.25 \times 10^{47}$ [1/s] particles of energy
$10^{18}$ [eV] is energetically allowed. The same source can emit
${\cal N}_{0} \simeq 6.25 \times {10}^{56}$ [1/s] particles of energy $10^{9}$ [eV] and so on.
If one follows \cite{Krauss} and assumes that for laboratory systems a typical
value of the ratio ${(E_{\phi}^{0}} - E_{min})/{{\it\Gamma}_{\phi}^{0}}$ is
$({E_{\phi}^{0}}- E_{min})/{{\it\Gamma}_{\phi}^{0}} \,
\geq \, O (10^{3} - 10^{6})$ and then taking, eg.
$({E_{\phi}^{0}}-E_{min})/{{\it\Gamma}_{\phi}^{0}} =
10^{6}$ one obtains from (\ref{t-as}) that
${\cal N}_{\phi}(T) \sim 2.53 \times 10^{-26}\, {\cal N}_{0}$
and from the estimation of $T$  used in \cite{Krauss} (see (11), (12) in \cite{Krauss})
that ${\cal N}_{\phi}(T) \sim   10^{-30}\,{\cal N}_{0} $. This means
that  there are ${\cal N}_{\phi}(T) \sim 14 \times 10^{21}$ particles per second of
energy $W^{\phi} = 10^{18}$ [eV] or ${\cal N}_{\phi}(T) \sim 14 \times 10^{30}$ particles
of energy $W^{\phi} = 10^{9}$ [eV] in the case of the considered example and $T$
calculated using (\ref{t-as}). On the other hand from $T$ obtained for the model
considered in \cite{Krauss} one finds ${\cal N}_{\phi}(T) \sim 6.25 \times  10^{17}$ and
${\cal N}_{\phi}(T) \sim 6.25 \times 10^{26}$ respectively. These estimations show
that astrophysical sources are able to create such numbers ${\cal N}_{0}$ of unstable
particles that sufficiently large number ${\cal N}_{\phi}(T) \gg 1$ of them has to survive up to
times $T$ when the effect described above should occur.
So the numbers of unstable
particles produced by some astrophysical sources are sufficiently large in order that
a significant part of them had to survive up to the transition times and therefore
to emit electromagnetic radiation.
The expected spectrum of this radiation can be very wide:
From radio frequencies up to $\gamma$--rays depending on energy distribution
function $\omega (E)$ of the unstable particle emitting this radiation.

\section{Final remarks}

We have shown that charged unstable particles or neutral unstable particles with
non--zero magnetic moment, which survived up to transition times or longer, should
emit electromagnetic radiation. We have also shown that only astrophysical processes
can generate sufficiently huge number  of unstable particles in order that this
emission could occur. From our  analysis it seems to be clear that  the effect described
in this paper may have an astrophysical meaning and help  explain
the controversies, which still remain, concerning the  mechanisms that generates the
cosmic microwave, or $X$--, or $\gamma$--rays emission,
e.g. it could help  explain why  some space areas (bubbles)  without visible  astronomical
objects  emit microwave radiation, $X$-- or $\gamma$--rays. Indeed,  let us consider
active galactic nuclei as an example.
They emit extremely huge numbers of stable and unstable particles including
neutrons (see eg. \cite{Anchordoqui})
along the axis of rotation of the galaxy.
The unstable
particles, which reached distances $d \sim d_{T}$ from the galactic plane,
should emit electromagnetic radiation.
So a distant observer should detect  enhanced emission of this radiation coming from
bubbles with the centra located  on the axis of the galactic rotation at average
distances $d_{T}$ from the galactic plane (see Fig. (\ref{f3})).
In the case of neutrons
$d_{T_{n}}$ can be extremely large.
Therefore a possible emission of the electromagnetic radiation  generated by
neutrons surviving sufficiently long seems to be relatively  easy to observe and
it should be possible to determine $d_{T_{n}}$.
Now having realistic sufficiently accurate $\omega(E)$ for neutrons we are able to calculate
$T_{n}$ and to find ${\cal E}_{n}(t)$
and its local maxima at transition times. Thus if the energies $W^{n}$,
(i.e., $\gamma_{L}$), are known then in fact we know velocities $v^{n}$
and we can compute $d_{T_{n}}$ and distances
where ${\cal E}_{n}(t)$ has  maxima.
All these distances fix the space areas where the
mechanism discussed
should manifest itself.
This suggests how to test this mechanism: The computed $d_{T_{n}}$
can be compared with observational data and thus one can test if the
mechanism described in our letter works in astrophysical processes.

Note that all possible  effects
discussed in this paper are the simple
consequence of the fact that the instantaneous energy
${\cal E}_{\phi}(t)$ of unstable particles becomes
large for suitably long times
compared with $E_{\phi}^{0}$   and for
some times even extremely large.
This property of ${\cal E}_{\phi}(t)$
is a purely quantum effect resulting
from the assumption that the energy spectrum is
bounded from below and it was found
by performing an analysis
of the properties of the quantum
mechanical survival probability $a(t)$.\\
\hfill\\

\noindent
{\bf Acknowledgments:} This work was supported in  part (KU)
 by the Polish NCN project 2013/09/B/ST2/03455.\\

\end{document}